\documentclass{llncs}
\usepackage{epsfig}
\usepackage{amsmath}
\usepackage{graphicx}
\usepackage{subfigure}
\usepackage{epstopdf}
\usepackage{subfig}
\usepackage{url}
\usepackage[stable]{footmisc}
\usepackage{setspace}
\title{A Unified Relevance Retrieval Model by Eliteness Hypothesis}
\author{Jagadeesh Gorla\inst{1}, Stephen Robertson\inst{2} \and Jun Wang\inst{1}\\
\email{\{j.gorla,j.wang\}@cs.ucl.ac.uk, stephenerobertson@hotmail.co.uk}
\institute{University College London \and Microsoft Research, Cambridge}}
\begin{document}
\maketitle
\begin{abstract}
 In this paper, an Eliteness Hypothesis for information retrieval is proposed, where we define two generative processes to create information items and queries. By assuming two types of deterministic relationships between the eliteness of terms and relevance, we obtain a new theoretical retrieval framework. The resulting ranking function is a 
unified one as it is capable of making use of  available relevance information on both the document and the query, which is otherwise unachievable by existing retrieval models. Our preliminary experiment on a simple ranking function has demonstrated the potential of the approach. 
\end{abstract}
\section{Introduction}
A central problem in information retrieval modeling is concerned with the estimation of the probability of relevance between documents and queries. In existing probabilistic retrieval models, the probability of relevance is calculated under two views. The first view is conditioned by the given document. That is to correlate a given document with the information need properties of the users who would judge it relevant. The second view is conditioned by the given information need (query). That is to correlate the given information need with the information properties of those documents that they would be judged relevant \cite{unified1}.  The former is called the \emph{document-oriented view}, including Maron and Kuhn's Probability Indexing and the language models \cite{Maron,Ponte98alanguage}; whereas the latter is called the \emph{query-oriented view}, represented by the Robertson-Sp\"{a}rck Jones model \cite{robertson:weight}. The two views rely on fixing one variable and optimising the other, i.e. conditioned by the information need and tuning the document or the other way around, but not both \cite{Microsoft03theunified}. Robertson et al \cite{unified1} first identified this problem and proposed an initial theoretical framework of unified model to unify both the views. One of the advantages of such modelling is that it is capable of using available relevance information on both the document and the query.

In this paper, we formulate a novel unified retrieval framework based on the ``Eliteness Hypothesis'' proposed in \cite{harter75probabilistic}. The framework is validated by implementing and testing a simple resulting ranking formula using the Expectation-Maximization parameter estimation method \cite{dempster77}.

The basic principle of a unified model is the following \cite{Microsoft03theunified}: firstly describe the documents and queries using a set of properties, including the vocabulary terms used to express them, and then the retrieval is performed by computing the relevance between document-query pairs as a function of description their properties. It should be possible to modify the descriptions of both the document and query properties once the relevance information has been available. Based on this idea, we define an Eliteness Hypothesis under the assumption that any given information items (document or query) can be described using a set of elite properties, and the relevance between a document-query pair is then dependent only on the elite properties and it can be thus estimated as a function of their elite properties. 

In the following sections, we first discuss the concept of eliteness (Section 2) then we state the ``Eliteness Hypothesis'' for IR and define two deterministic relevance relations between a document-query 
pair, namely ``Strict Identity'' and   ``Logical inclusion'', under the hypothesis (Section 3). In section 4, we derive a unified probabilistic relevance ranking function for IR based on Eliteness Hypothesis 
by probabilistically modelinBased on thisg the eliteness of elite propeRelevance plays a critical role in information retrieval modelling. In many mathematical formulations, the probability of relevance is calculated either by correlating the documents with the information need properties of the users who would judge them relevant, or by correlating the users information needs with the information properties of those documents that they would be judged relevant. In other words, existing probabilistic retrieval models are capable of using the relevance information present either on query or document but not both. In this paper, we propose a theoretical framework to develop a unified probabilistic relevance model. The intention is to make use of the available relevance information on both the query and the document in order to compute the probability of relevance between them. Our development is based on the "Eliteness hypothesis", which assumes that any document or a query can be described using a set of elite properties. The relevance between the document and query is dependent only their elite properties and these properties can be modified in the presence of relevance feedback. We derive probabilistic relevance ranking functions based on our hypothesis and as a preliminary study, evaluate one of simplified versions on the TREC-8 ad hoc task collection.rties. Furthermore, we derive two probabilistic ranking functions from unified relevance ranking function using ``Strict identity'' and  
``Logical inclusion'' relationships along with some valid simplifying assumptions. For implementing a basic ranking function, in order to validate the proposed theoretical framework, we present an 
estimation methode for estimating the eliteness of elite properties by assuming the document collection is a 2-Poisson mixture \cite{RobertsonW:1994} of elite and non-elite documents for a given elite 
property (Section 5). Finally, we present experimental results on the basic ranking function and show that it performs better than other influential ranking functions on TREC-8 ad hoc collection. 
We will also present some discussion about the basic ranking function relationship with the Inverse Document Frequency \cite{Jones72astatistical}. 
\section{Eliteness}
The idea of ``eliteness'' is to describe a set of elite properties (or concepts) represented by an information. 
In general, in the context of IR, we can classify elite properties into two different sets of properties: 
\begin{enumerate}
 \item properties that describe both information item (document) and information need (query). A typical example of these properties are the properties associated with the terms in vocabulary. 
  These properties could be the natural language phrases used to express the information but in this paper we use only the elite properties that are associated with each term in the vocabulary. 
  We name the properties associated with the single term properties as term-description properties.  \item properties that describe only document (e.g. page rank, url depth, etc) or only query.  \end{enumerate} We define the eliteness of an elite property for the given information, on the basis of the definition of eliteness introduced in \cite{RobertsonW:1994,harter75probabilistic} as: \begin{itemize}
\item the eliteness value of an elite property for the given information is binary, i.e. it is either elite (1) or non-elite (0). We can also say that the eliteness of a piece of information to a given elite property is a binary one.
 \item If the piece of information is \emph{about} an elite property then its eliteness is elite for the property otherwise non-elite.  
\end{itemize}

From the definition of eliteness, for any given elite property, some documents in the collection are ``elite'' and others are ``non-elite''. Inspired by a 2-Poisson model proposed in \cite{harter75probabilistic,RobertsonW:1994}, we state our Eliteness Hypothesis as:
\vspace{12pt}

\emph{``Any information, expressed by an author or a user, can be described using a binary set of elite properties. If we know the eliteness value of each elite property for a given information item 
(document) and need (query), we can deterministically determine whether the item-need pair is relevant or not.''}

\vspace{12pt}

We further assume that an author (or a user) will carry out the following process to express their information:
\begin{enumerate}
\item First, a user or author will choose a set of elite properties such that these properties can describe every aspect of the information that they want to express. The chosen elite properties are ``elite'' and the rest are ``non-elite''.
\item Once the elite properties are chosen, an observable information item or need, is generated by a stochastic function of chosen elite properties. The uncertainty about the eliteness 
is injected during the generation process.  
\end{enumerate}

Our hypothesis suggests that the relevance between document $d$ and query $q$ is known exactly if their elite properties, represented by $\textbf{E}$, $\textbf{F}$ respectively, are observed. 
Two deterministic relations are defined as follows:

A document $d$ and query $q$ are relevant if 
\begin{itemize}
\item Strict identity: $\textbf{E} =\textbf{F}$, i.e. the eliteness of all the elite properties of $d$ must be the same as the eliteness of the elite properties of $q$, or
\item Logical inclusion: $\textbf{F} \subset \textbf{E}$, i.e. the elite properties of the document with eliteness ``elite'' must contain all the elite properties of the query whose eliteness 
is ``elite''. 
\end{itemize}

The other possibilities of defining the relevance under the hypothesis are stochastic rather than deterministic but would introduce an additional stochastic layer into the model. For simplicity, this paper only considers the deterministic relation and leave the stochastic one for future work. However, it should be noticed that the relevance between $d$ and $q$ will still be a probabilistic function given the generative process from Step 2 and the fact that the eliteness is not observable.  The unification is achieved by modifying the eliteness probabilities of the document-query pair when the relevance information is available on the document and/or the query, which is otherwise not provided by existing retrieval models.

\section{The Probabilisti
c Relevance Ranking Function}
We are ready to derive a probabilistic ranking function. Let $D$ be a random variable whose value is any possible document  generated by an author. Similarly, Let $Q$ be a random variable whose 
possible values are any possible query generated by a user. We use the lower case $d$ and $q$ to denote their particular instantiations respectively.  Let $\textbf{E} \in \{0,1\}^k$ be a random 
binary vector over the space defined by the document's elite properties, where $k$ is the number of elite properties, e.g. one elite property for each word in vocabulary.  
Similarly, $\textbf{F}\in \{0,1\}^k$ is a random binary vector over the space defined by the query's elite properties. Their elements are denoted as ${E}_{i}$ and ${F}_{i}$ respectively, 
where $i\in\{1,k\}$. $E_i= 1$ means ``elite'', whereas $E_i= 0$ means ``non-elite''.

Our eliteness hypothesis gives the probability of relevance $P(R=1 | d, q)$ as
\begin{equation}\label{eq1}
P(R=1 | d, q)  = \sum_{\textbf{E}} \sum_{\textbf{F}} P(R=1, \textbf{E}, \textbf{F} | d, q )
\end{equation} 
where $R$ is a binary random variable. $R=1$ means relevant while $R=0$ otherwise. Applying Bayes' rule gives
\begin{equation}\label{eq2}
P(R=1 | d, q)  = \sum_{\textbf{E}} \sum_{\textbf{F}} \underbrace{ P(R=1 |\textbf{E}, \textbf{F} ,d, q ) }_\text{part one} \underbrace{P(\textbf{E}, \textbf{F}| d, q)}_\text{part 2}
\end{equation}

From the hypothesis, $\textbf{E}$ is independent of $q$ and $\textbf{F}$, $\textbf{F}$ is independent of $d$ and $\textbf{E}$ and \textbf{E}, \textbf{F} are sufficient to determine the relevance between $d$, $q$. Therefore, by ignoring $d$ and $q$ and applying independence assumptions to Eq.~\eqref{eq2}, we get

\begin{equation}\label{eq3} P(R=1 | d, q) = \sum_{\textbf{E}} \sum_{\textbf{F}} \underbrace{P(R=1| \textbf{E}, \textbf{F} )}_\text{part 1} \underbrace{P(\textbf{E}| d) P(\textbf{F}| q)}_\text{part 2} \end{equation} 

By applying Bayes' rule to Eq.~\eqref{eq3} and removing constant $P(R=1)$  we get 

\begin{equation}\label{eq5} P(R=1 | d, q) \propto_{R}  \sum_{\textbf{E}} \sum_{\textbf{F}} \underbrace{\frac{ P(\textbf{E}, \textbf{F}| R=1) } { P(\textbf{E}) P(\textbf{F})}}_\text{part 1} \underbrace{P(\textbf{E}| d) P(\textbf{F}| q)}_\text{part 2} \end{equation}
where $\propto_{R}$ denotes the rank equivalent. Here, we make an assumption, similar to term-independence assumption in \cite{robertson:weight}, that the eliteness of an elite property is independent of other elite properties.  We thus have
\begin{equation} \label{eq6}
P(R=1 | d, q)  \propto_{R} \sum_{\textbf{E}} \sum_{\textbf{F}} \prod_{i}^{k} \Big( \frac{ P(E_{i}, F_{i}| R=1) } { P(E_{i}) P(F_{i})}  P(E_{i}| d) P(F_{i}| q)\Big)
\end{equation}
The Eq.~\eqref{eq6} is a unified probabilistic relevance ranking function.  It uses the information about each elite property between the document and query in the relevant set ($P(E_{i}, F_{i}| R=1)$). Thus, the information about other relevant document-query pairs that share the eliteness of elite property is included, which is an essential component of a unified model \cite{unified1}.

In order to test the theory, we make the following assumption called \emph{Query Eliteness Assumption} and derive relevance ranking functions from unified ranking function using the Strict Identity and Logical Inclusion relations.

\emph{Query Eliteness Assumption:} We assume that the eliteness of each elite property is known for a given query, i.e. we know the binary vector \textbf{F}.  This assumption is similar to an implicit assumption made the query terms are elite to the query and other terms are non-elite in \cite{RobertsonW:1994}.  Let $fq_{i}$ denotes the eliteness value of $F_{i}$ for each of term, namely $F_{i} = fq_{i}$ where $fq_{i}=1$ if $F_{i}$ is elite for $q$ otherwise $fq_{i}=0$.
Based on the our assumption, Eq.~\eqref{eq6} can be written as 
\begin{equation}\label{eq7}
P(R=1 | d, q) \propto_{R} \sum_{\textbf{E}}  \prod_{i=1}^{k} \Big( \underbrace{\frac{ P(E_{i}, F_{i}=fq_{i}| R=1) } { P(E_{i}) P(F_{i}=fq_{i})}}_\text{part 1} P(E_{i}| d) P(F_{i}=fq_{i}| q)\Big)
\end{equation}
By applying Bayes' rule to part 1 of the Eq.~\eqref{eq7}, we get 
\begin{multline}\label{eq8}
P(R=1 | d, q) \\ \propto_{R} \sum_{\textbf{E}}  \prod_{i=1}^{k} \Big( \frac{ P(E_{i}| F_{i}=fq_{i}, R=1) P(F_{i}=fq_{i}| R=1)} { P(E_{i}) P(F_{i}=fq_{i})} 
 P(E_{i}| d) P(F_{i}=fq_{i}| q)\Big)
\end{multline}
Further factorizing Eq.~\eqref{eq8} gives
\begin{multline}\label{eq9}
 P(R=1 | d, q) \\\propto_{R} \sum_{\textbf{E}}  \prod_{\forall i : F_{i} = 0} \Big( \frac{ P(E_{i}| F_{i}=0, R=1) P(F_{i}=0| R=1)}{P(E_{i})P(F_{i}=0)}  P(E_{i}| d) P(F_{i}=0| q)\Big) \\
 \prod_{\forall i:F_{i}=1} \Big( \frac{ P(E_{i}| F_{i}=1, R=1) P(F_{i}=1| R=1)} { P(E_{i}) P(F_{i}=1)} P(E_{i}| d) P(F_{i}=1| q)\Big) \end{multline} 
From the Query Eliteness Assumption, we know that the value of each element in \textbf{F}. That is there is no uncertainty in the eliteness of elite properties for the given $q$. So we have $P(F_{i}=1| q) =1$ if term $i$ is in query $q$ and $P(F_{i}=0| q) =1$ otherwise. By substituting the above values, Eq.~\eqref{eq9} becomes \begin{multline}\label{eq10}
 P(R=1 | d, q)  \\\propto_{R}\sum_{\textbf{E}}  \prod_{\forall i:F_{i} = 0} \Big( \frac{ P(E_{i}| F_{i}=0, R=1) P(F_{i}=0| R=1)} { P(E_{i}) P(F_{i}=0)}  P(E_{i}| d)\Big) \\ \prod_{\forall i :F_{i} = 1} \Big( \frac{ P(E_{i}| F_{i}=1, R=1) P(F_{i}=1| R=1)} { P(E_{i}) P(F_{i}=1)}  P(E_{i}| d)\Big)
\end{multline}
From here onwards, we use  Eq.~\eqref{eq10} to derive a relevance ranking functions equivalent of the relevance based on Strict identity and  Logical inclusion.  
\subsection{Ranking function under Strict identity}
As defined under the Strict Identity relation, a document and a query is relevant if \textbf{F} $=$ \textbf{E}.  From the definition, we can assume that if we know that the eliteness of an elite property is ``elite'' for the given query (i.e. $F_{i}=1$), then the probability that eliteness of same property 
  is ``elite'' ($E_{i}=1$) in the relevant set of documents, i.e. $P(E_{i} =1| R=1, F_{i}=1) = 1$, as they have the same eliteness value. Equally, it is the same for the non-elite case where $P(E_{i} =0| R=1, F_{i}=0) = 1$. Note that from this assumption, the score of any vector in $\textbf{E}$ of Eq.~\eqref{eq10} is zero if $E_{i}=0$ and $F_{i} = 1$ (or) $E_{i}=1$ and $F_{i} = 0$ for 
at least one $i$.
By substituting the above values in Eq.~\eqref{eq10}, we get 
\begin{multline} \label{eq11}
 P(R=1 | d, q)   \propto_{R}\prod_{\forall i=1:F_{i} = 0} \Big( \frac{ P(F_{i}=0| R=1)} { P(E_{i}=0) P(F_{i}=0)}  P(E_{i}=0| d)\Big) \\
\prod_{\forall i:F_{i} = 1} \Big( \frac{ P(F_{i}=1| R=1)} { P(E_{i}=1) P(F_{i}=1)}  P(E_{i}=1| d)\Big)
\end{multline}
where $P(F_{i}=0| R=1)$, $ P(F_{i}=1)$, $P(F_{i}=0)$ and $P(F_{i}=1| R=1)$ in Eq.~\eqref{eq11} can be removed as these terms do not effect the ranking order. We thus get 
\begin{equation}\label{eq12}
P(R=1 | d, q)  \propto_{R} \prod_{\forall i=1 :F_{i} = 0} \Big( \frac{P(E_{i}=0| d)} {P(E_{i}=0) } \Big)  \prod_{\forall i:F_{i} = 1} \Big( \frac{ P(E_{i}=1| d) } {P(E_{i}=1)}  \Big)
\end{equation}
Eq.~\eqref{eq12} is a ranking function under the Strict identity relation with the Query Eliteness Assumption.
\subsection{Ranking function under Logical inclusion}
As defined under the Logical Inclusion relation, a document is relevant to a query if and only if \textbf{F} $\subset$ \textbf{E}.  We can thus assume $P(E_{i} =1| R=1, F_{i}=1) = 1$ as all the elite properties with eliteness value ``elite'' for a query must have the same eliteness value in a relevant set of documents. 
That is, if a property $F_{i} =1$ then $E_{i}=1$ in relevant documents. By applying the above assumption, Eq.~\eqref{eq10} becomes
\begin{multline}\label{eq13}
P(R=1 | d, q)  \propto_{R} \underbrace{\sum_{\textbf{E}^{'}}  \prod_{\forall i:F_{i} = 0} \Big( \frac{ P(E_{i}| F_{i}=0, R=1) P(F_{i}=0| R=1)} { P(E_{i}) P(F_{i}=0)} P(E_{i}| d)\Big)}_\text{part 1} \\
\underbrace{\prod_{\forall i:F_{i} = 1} \Big( \frac{P(F_{i}=1| R=1)} { P(E_{i}=1) P(F_{i}=1)} P(E_{i}=1| d)\Big)}_\text{part 2}
\end{multline}
where $\textbf{E}^{'}$ is the set of all possible vectors with $E_{i}=1 ~\forall F_{i}=1$. Part 2 of the Eq.~\eqref{eq13} is same for all the vectors in $\textbf{E}^{'}$.
So, we can rewrite Eq.~\eqref{eq13} as 
\begin{multline}\label{eq14}
  P(R=1 | d, q)  \propto_{R}\prod_{\forall i :F_{i} = 1} \Big( \frac{ P(F_{i}=1| R=1) P(E_{i}=1| d)} { P(E_{i}=1) P(F_{i}=1)} \Big) \\ \Big( \sum_{\textbf{E}^{'}}
  \prod_{\forall i:F_{i} = 0} \Big( \frac{ P(E_{i}| F_{i}=0, R=1) P(F_{i}=0| R=1)} { P(E_{i}) P(F_{i}=0)} P(E_{i}| d)\Big)\Big)
\end{multline}
Eq.~\eqref{eq14} is a relevance ranking function under Logical Inclusion with \emph{Query eliteness assumption}.

We can further simplify the Eq.~\eqref{eq14} by ignoring the terms that are rank independent and the terms with non-elite properties of the query in Eq.~\eqref{eq14}. This is similar to ignoring the terms that are not present in query in \cite{robertson:weight,Ponte98alanguage,zl}. Applying a logarithm transform results in the following ranking function:
 \begin{equation}\label{eq15} P(R=1 | d, q) \propto \sum_{\forall i:F_{i} = 1} \log \frac{ P(E_{i}=1| d) } {P(E_{i}=1)}
 \end{equation}
 \subsubsection{Relationship with IDF:} One of the interesting by-products is that the above formula provides a yet anther theoretical justification of IDF (inverse document frequency) as scoring function \cite{tfidf}. To see this, let us assume that the eliteness of an term-description elite property is elite to a document if the term is present in the document and non-elite otherwise.  Then, the probability of the eliteness of elite property in the collection is to be $P(E_i=1):= \frac{n_{ci}}{N}$, where $n_{ci}$ is the number of documents in the collection with the term associated with the $i^{th}$ term-description property and $N$ is the total number of documents in the collection. From the above assumption, $\mu_i(0) = 0$; $P(E_i|d):=1$ if $F_i=1$ (term $i$ is in the query). By substituting them in the ranking function Eq.~\eqref{eq15} we get
 \begin{equation}\label{idf}
 P(R=1 | d, q)  \propto \sum_{\forall i:F_{i} = 1} \log \frac{1}{p_i} =  \sum_{\forall i:F_{i} = 1} \log  \frac{N}{n_{ci}} 
\end{equation} 
The ranking function in Eq.~\eqref{idf} is simply a function of IDF values of the query terms. Basically it says that the IDF score function relies on the assumption that a term is elite if it occurs in the document.
This is different from the explanation provided by the Robertson-Sp\"{a}rck Jones model, where an explicit assumption that the whole collection is non-relevant set is needed \cite{robertson:weight}.
\section{A Simple Ranking Function and its Experiments} 
In order to test the theory, we have implemented the basic ranking function in Eq.~\eqref{eq15}.  To compute Eq.\eqref{eq15}, we first define a single elite property for each term in the vocabulary and use this set of properties as a complete set of elite properties.

\subsection{Estimating Eliteness Probabilities}
From the Eliteness Hypothesis, we know that, the occurrence of a term in a document has a stochastic element associated with the eliteness of its corresponding term-description property.
Therefore, we compute the probability of $i^{th}$ term-description property being ``elite'' for the given document $d$ as  
\begin{equation} \label{eq16}
 P(E_{i}=1| d) \equiv P(E_{i}=1| tf_i)
\end{equation}
 where $tf_i$ denotes the term frequency associated with $i^{th}$ term-description property in document $d$. From the hypothesis, some documents in the collection are 
``elite'' to an elite property and others are non-elite. And, $tf_i$ follows one distribution in elite set of documents and another distribution in non-elite set of documents.
Therefore, we can draw a probabilistic inference about the eliteness of a term-description property from its associated term's term frequency in the document.

By applying Bayes' rule to $P(E_{i}=1| tf_i)$, we get 
\begin{equation} \label{eq171}
 P(E_{i}=1| tf_i) = \frac{P(tf_i|E_{i}=1) P(E_{i}=1)}{P(tf_i)} =  \frac{P(tf_i|E_{i}=1) P(E_{i}=1)}{\sum_{E_{i}\in\{0,1\}}P(tf_i|E_{i}) P(E_{i})}.
\end{equation}

By using Eq.~\eqref{eq171} we compute ranking function value for each property in Eq.~\eqref{eq15} as follows
\begin{equation} \label{eq17}
 \frac{P(E_{i}=1| tf_i)}{ P(E_{i}=1)} = \frac{P(tf_i|E_{i}=1)}{\sum_{E_{i}\in\{0,1\}}P(tf_i|E_{i}) P(E_{i})}.
\end{equation}

For simplicity, we consider only term-description properties, and use query terms to represent elite properties of the query ($F_{i} = 1$ when $q_{i} = 1$).  By substituting Eq.~\eqref{eq17} back in Eq.~\eqref{eq15}, we get the following ranking function
\begin{multline}\label{eq18} 
  P(R=1 | d, q) \\\propto \sum_{\forall i:F_{i} = 1} \log \frac{ P(E_{i}=1| d) } {P(E_{i}=1)}
  =\sum_{\forall i:F_{i} = 1} \log  \frac{P(tf_i|E_{i}=1)}{\sum_{E_{i}\in\{0,1\}}P(tf_i|E_{i}) P(E_{i}) }
\end{multline} To compute Eq.~\eqref{eq18}, we adopt the 2-Poisson model in \cite{harter75probabilistic,RobertsonRP80}.  

\subsection{The 2-Poisson mixture and the EM algorithm} According to the 2-Poisson model hypothesis \cite{harter75probabilistic,RobertsonRP80}, if the elite property is a term-description property then its associated term's frequency follows a Poisson distribution in the elite set of documents \begin{equation} P(tf_i|E_{i}=1) := e^{-\mu_{i}(1)} \mu_{i}(1)^{tf_i} \end{equation} and another Poisson distribution in the non-elite set \begin{equation} P(tf_i|E_{i}=0) := e^{-\mu_{i}(0)} \mu_{i}(0)^{tf_i} \end{equation} where $\mu_{i}(1)$ and $\mu_{i}(0)$ are the two Poisson means.  The mixing probability $P(E_i=1):=p_i$ is an additional parameter.  The 2-Poisson mixture can be written as \begin{align}\label{eq19}
  \nonumber P(tf_{i}| \theta) = &  \sum_{E_{i}\in\{0,1\}}P(tf_i|E_{i}) P(E_{i}) \\
  = & p_i \frac{e^{-\mu_{i}(1)} \mu_{i}(1)^{tf_i}}{tf_{i}!} + (1-p_i) \frac{e^{-\mu_{i}(0)} \mu_{i}(0)^{tf_i}}{tf_{i}!}  \end{align} where $\theta$ is unknown parameters of the mixture and $ \theta = (\mu_{i}(1), \mu_{i}(0), p_i)$. Different with \cite{harter75probabilistic,RobertsonRP80}, we obtained the optimal maximum likelihood parameters using the Expectation Maximization algorithm \cite{dempster77}, where the likelihood function is \begin{align}\label{eq20}
  \nonumber \Lambda(tf_i;\theta) =& \prod_{n=1}^{N} P(tf_{i}^{n}|\theta)  \\
  = &\prod_{n=1}^{N} \Big( p_i \frac{e^{-\mu_{i}(1)} \mu_{i}(1)^{tf_i^n}}{tf_{i}^n!} + (1-p_i) \frac{e^{-\mu_{i}(0)} \mu_{i}(0)^{tf_i^n}}{tf_{i}^n!}\Big) \end{align} 
where $N$ is the total number of documents and $tf_{i}^{n}$ denotes the term frequency of the term associated with $i^{th}$ term-description elite property in $n^{th}$ document ($d_n$) of the collection.  

The parametric estimation problem can be defined as 
\begin{align} \hat{\theta} = \operatorname*{arg\,max}_{\theta} \Lambda(tf_i;\theta) \end{align} 
where the parameters can be obtained by maximizing the log-likelihood function $\Lambda$. More specially, we compute the partial derivatives of the logarithm of the likelihood function with respect to $\mu_{i}(1)$, $\mu_{i}(0)$ and $p_i$, and setting them to zero. The logarithm of the likelihood function $\Lambda(tf_i;\theta)$ is \begin{equation}
 \lambda(tf_i;\theta ) :=  \sum_{n=1}^{N} \log \Big( p_i \frac{e^{-\mu_{i}(1)} \mu_{i}(1)^{tf_i^n}}{tf_{i}^n!} +  (1-p_i) \frac{e^{-\mu_{i}(0)} \mu_{i}(0)^{tf_i^n}}{tf_{i}^n!}  \Big)
\end{equation}
By setting
\begin{equation}
 \frac{\partial \lambda }{\partial \mu_{i}(1)} = 0, \frac{\partial \lambda }{\partial \mu_{i}(0)} = 0
\end{equation} 
 we get,
\begin{equation}
\mu_{i}(1) = \frac{ \sum_{n=1}^{N} P(E_{i} = 1|tf_i^n) ~ tf_i^n}{\sum_{n=1}^{N} P(E_{i} = 1 |tf_i^{n})}
\end{equation} and 
\begin{equation}
\mu_{i}(0) = \frac{ \sum_{n=1}^{N} P(E_{i} = 0|tf_i^n) ~ tf_i^n}{\sum_{n=1}^{N} P(E_{i} = 0 |tf_i^{n})}.
\end{equation}
By setting
 \begin{equation}
 \frac{\partial \lambda }{\partial p_{i}} = 0
 \end{equation}
 we get, \begin{equation} p_i = \frac{1}{N} \sum_{n=1}^{N} P(E_{i} = 1| tf_i^{n}) \end{equation} 
where $P(E_{i}=1| tf_i^n)$ can be computed using Eq.~\eqref{eq171}. 

In order to find the optimal parameters, we start with an initial estimates of $p_{i}^{(0)}$, $\mu_{i}(1)^{(0)}$, and $\mu_{i}(0)^{(0)}$, and then the better estimates can be found by iterating through the following E- and M-step \cite{dempster77}.  
\begin{enumerate}
 \item E-step:
 \begin{equation}
 P^{(k+1)} (E_{i}=1|tf_{i}) = \frac{P^{(k)}(tf_i| E_{i}=1) p_i^{(k)}}{\sum_{E_{i}\in\{0,1\}}P^{(k)}(tf_i|E_{i}) P(E_{i})}.
 \end{equation}
 \item M-step:
   \begin{equation} \mu_{i}(0)^{(k+1)} = \frac{ \sum_{n=1}^{N} P^{(k)}(E_{i}=0|tf_i^{n}) tf_i^{n}}{\sum_{n=1}^{N} P^{(k)}(E_{i}=0|tf_i^{n})}.  \end{equation}

\begin{equation} \mu_{i}(1)^{(k+1)} = \frac{ \sum_{n=1}^{N} P^{(k)}(E_{i}=1|tf_i^{n}) tf_i^{n}}{\sum_{n=1}^{N} P^{(k)}(E_{i}=1|tf_i^{n})}.  \end{equation}

\begin{equation} p_{i}^{(k+1)} = \frac{1}{N} \sum_{n=1}^{N} P^{(k)}( E_{i}=1|tf_i^{n}).  \end{equation} 
\end{enumerate} where $k$ is the step number. We measure the difference of the perplexity between two consecutive models of the probability distribution of $P(E_{i}=1|tf_i)$ to decide when shall 
we stop the training. By substituting the estimated parameter values in Eq.~\eqref{eq15} we reach the final ranking function \begin{equation}\label{strict identity}
   P(R=1 | d, q)  \propto \sum_{\forall i:F_{i} = 1} \log \Big( \frac{e^{-\mu_{i}(1)} \mu_{i}(1)^{tf_i}} {p_{i} e^{- \mu_{i}(1)} \mu_{i}(1)^{tf_i} +  (1-p_{i}) e^{-\mu_{i}(0)} \mu_{i}(0)^{tf_i} }\Big)
 \end{equation} 
 \subsection{Document Length Normalization}
 One of the issues in the 2-Poisson model is that it assumes a fixed document length for all the documents \cite{harter75probabilistic,RobertsonW:1994}. 
Generally, it is not a valid assumption. Two hypothesis were proposed to explain the varied document lengths in the collection namely Scope Hypothesis and Verbosity hypothesis \cite{RobertsonW:1994,Singhal}.
  Similarly, in this paper we assume the verbosity hypotheses, i.e. lengthy documents covers the same scope of short document but uses more words, and incorporate it into our ranking function by modifying
 the term frequency as 
 \begin{equation}
   tf = tf  \Big(b + (1-b) \frac{avgDL}{DL}\Big), 
\end{equation}
where $avgDL$ is average document length in the collection, $DL$ is the document length and $b\in[0,1]$. The above formula penalizes the term frequency in lengthy documents based on the  
 $b$ value.
\subsection{Experiments}
For the experiment, we employed the TREC-8 ad hoc task collection and topics to evaluate the ranking function in Eq.~\eqref{strict identity} against Mean Average Precision (MAP), Mean Reciprocal Rank (MRR) and Recall@1000 metrics. We initialized the parameters with the data collection statistics. The initial value of $p_{i}$ was set to be the portion of documents collection with the term (same as the assumption used to show the ranking function relationship with IDF), and $\mu_{i}(1)$ was initialized with the average number of times the term appeared in document with its term frequency more than one.  We used a minuscule value to initialize $\mu_{i}(0)$.  The assumption is that the average term frequency of a term associated with the term-description elite property in a document approaches zero if it is ``non-elite'' to the document.
  \begin{table}[t]
  \scriptsize
  \setlength{\abovecaptionskip}{2pt}
\centering
    \begin{tabular}{ | l | l | l | l | }
    \hline
    Model  & MAP & MRR & Recall@1000 \\ \hline
    BM25 &    0.250	&     0.638	&     0.6634  \\ \hline
    Language Model with JM smoothing &    0.238	&     0.4816	&     0.658 \\ \hline
    Language Model with Dirichlet prior &    0.2539   &      0.6376  &   0.6694 \\ \hline
    Unified Model &    0.2553 (0.2266$^{*}$)   &      0.607 (0.6513$^{*}$)  &    0.6659  \\ \hline
  \end{tabular}
   \caption{Performance  on the TREC-8 ad hoc task data collection. }
\label{table:performace}
\end{table}

\begin{figure*}[t] \begin{center} \begin{scriptsize} \begin{tabular}{ccc}
    \includegraphics[width=1.6in]{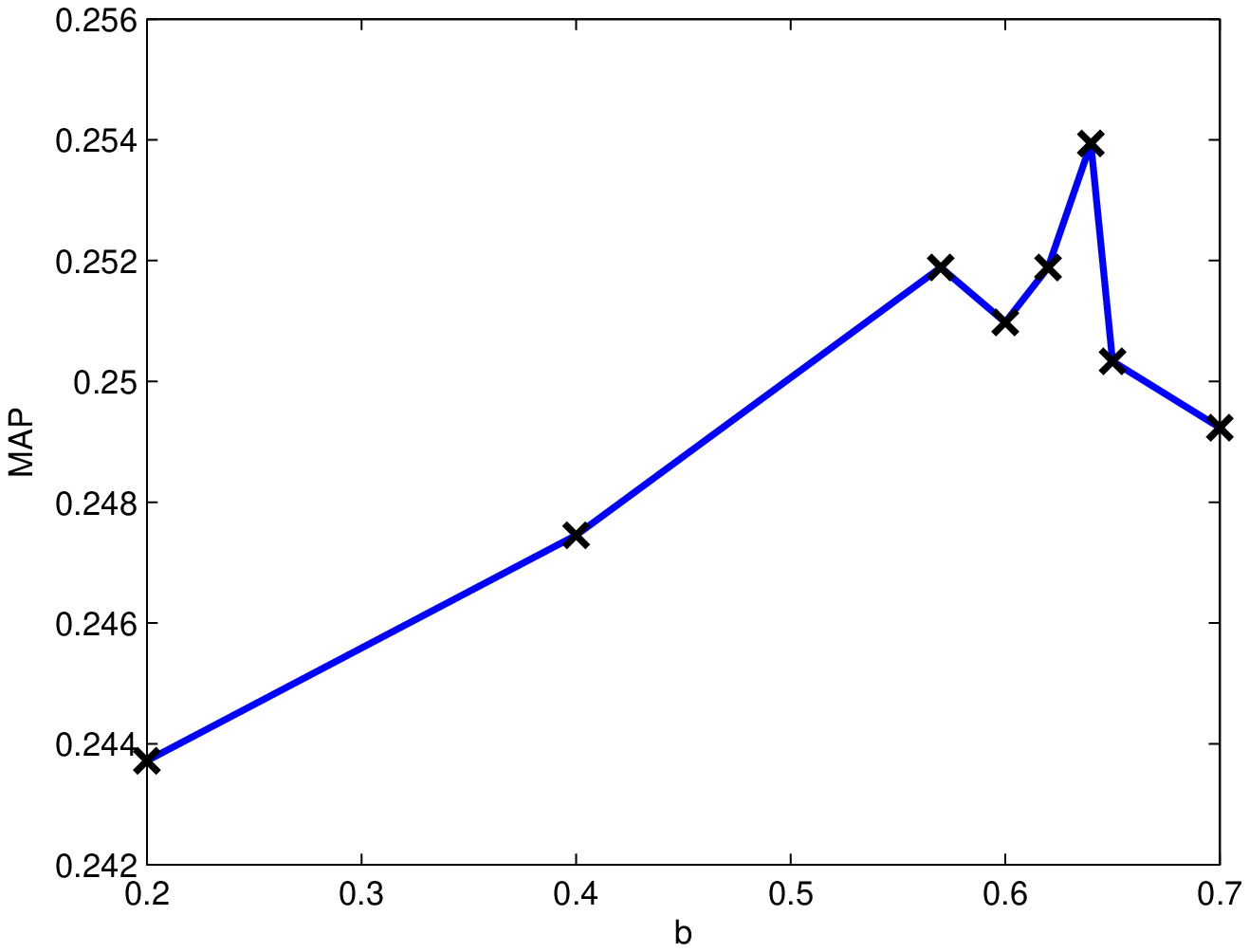}
&
    \includegraphics[width=1.6in]{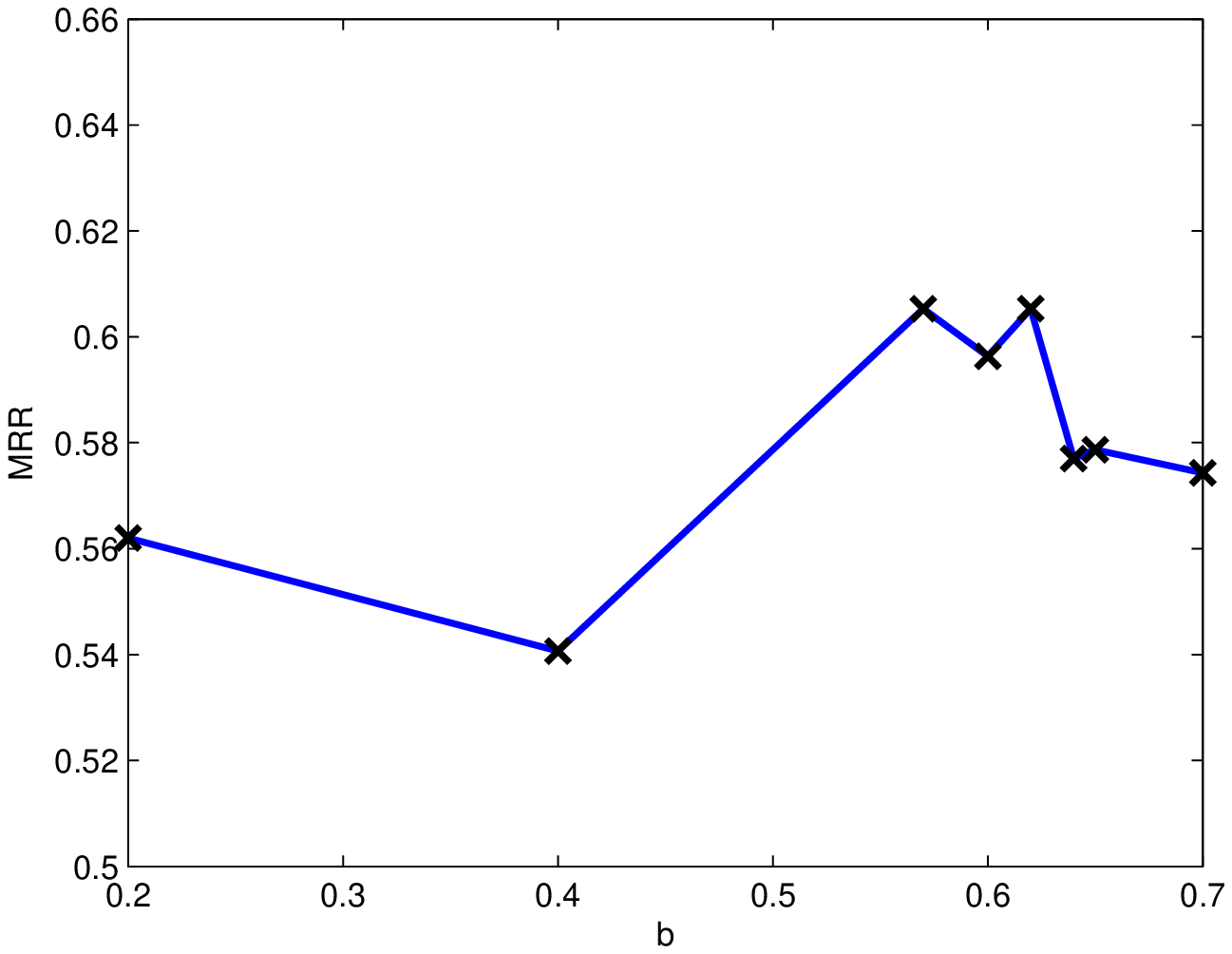}
    &
\includegraphics[width=1.6in]{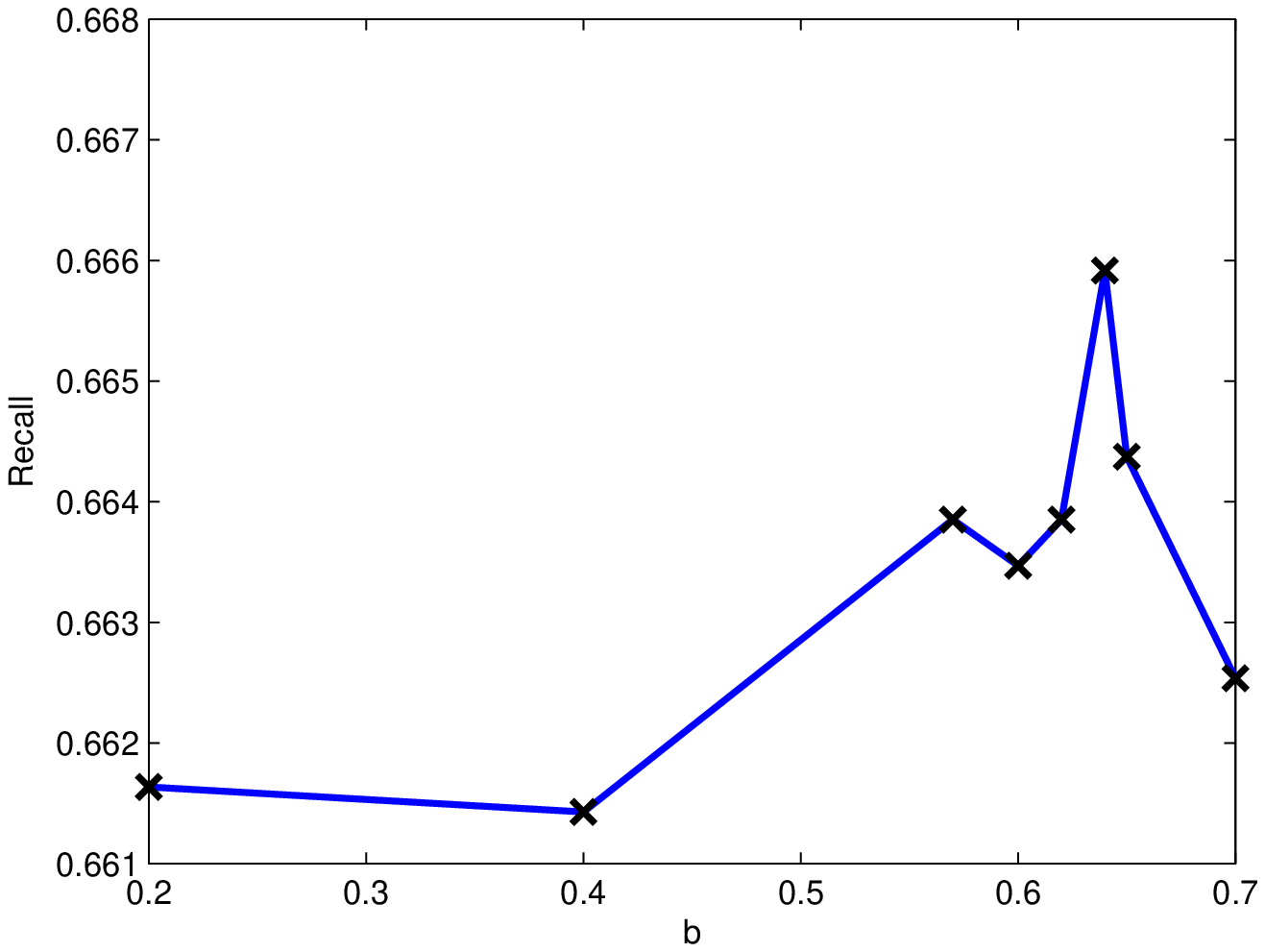}
\\
(a) MAP
&
(b) MRR
&
(c) Recall
\\
\end{tabular}
\end{scriptsize}
\end{center}
\vspace{-15pt}
\caption{Different metrics performances for different values of $b$ for a fixed value of $n=3$.}
\label{fig:valueb}
\end{figure*}

We have run tests and compared the performance of our method with BM25 \cite{Robertson:2009} and Language Models \cite{Ponte98alanguage} (with the Jensen-Mercer smoothing and Dirichlet prior).  Table \ref{table:performace} shows that our method outperforms the baselines when directly tuning the parameters with respect to the metrics. The improvement over MRR is significant. We also found that MAP and MRR cannot be optimized simultaneously and this confirmed the theoretical argument about the trade-off between MAP and MRR in \cite{irmetric}. 

As discussed, the parameter values were learned by boosting the initial value of $\mu_{i}(1)$ (i.e. the average term frequency in elite set of documents.) and by multiplying with an integer $n$, i.e. $\mu_{i}(1)^{(0)} = n.\mu_{i}(1)^{(0)}$.  To  see the effect of the document length normalization as well as the initialization of $\mu_i(1)$, we fixing the $n$, and vary $b$ and vice-versa. The graph in Fig.~\ref{fig:valueb} shows that the document length normalization indeed has an effect on performance and the best performance was achieved at $b=0.64$ for $n=3$.  

\section{Conclusion and Future work} 
We have proposed a theoretical unified retrieval framework based on the Eliteness Hypothesis and derived a unified probabilistic relevance ranking function. 
Our initial experiments with a basic ranking function demonstrated that the approach is indeed encouraging.  One of the problems with our current estimation is that the 
EM algorithm does not always converge to the global maximum. In the future, we would like to test the model by estimating the eliteness probabilities with Bayesian methods and 
 also remove \text{Query eliteness assumption} to incorporate the uncertainty in the eliteness of query elite properties and test the model.

\begin{scriptsize}
\bibliographystyle{abbrv}
\bibliography{sigproc}
\end{scriptsize}
\end{document}